# XUV Induced Bleaching of a Tin Oxo Cage Photoresist Studied by High Harmonic Absorption Spectroscopy

**Najmeh Sadegh[1], Maarten van der Geest[1], Jarich Haitjema[1], Filippo Campi[1], Sonia Castellanos[1], Peter M. Kraus[1*], and Albert M. Brouwer[1,2*]**

[1]*Advanced Research Center for Nanolithography, P.O. Box 93019, 1090 BA Amsterdam, The Netherlands*

[2]*University of Amsterdam, van 't Hoff Institute for Molecular Sciences, P.O. Box 94157, 1090 GD Amsterdam, The Netherlands*

\* *f.brouwer@arcnl.nl*
\* *p.kraus@arcnl.nl*

Inorganic molecular materials such as tin oxo cages are a promising generation of photoresists compatible with the demands of the recently developed Extreme UltraViolet (EUV) lithography technology. Therefore, a detailed understanding of the photon-induced reactions which occur in photoresists after exposure is important. We used XUV broadband laser pulses in the range of 25 – 40 eV from a table-top high-harmonic source to expose thin films of the tin oxo cage resist to shed light on some of the photo-induced chemistry via XUV absorption spectroscopy. During the exposure, the transmitted spectra were recorded and a noticeable absorbance decrease was observed in the resist. Dill parameters were extracted to quantify the XUV induced conversion and compared to EUV exposure results at 92 eV. Based on the absorption changes, we estimate that approximately 60% of tin-carbon bonds are cleaved at the end of the exposure.

## 1. Introduction

Metal containing molecular inorganic resists are considered as promising materials for Extreme Ultraviolet (EUV) lithography, and their development is crucial for the success of this technology[1]. Compared to the traditional organic resists the metal atoms in the inorganic cores provide high etch resistance and high absorption cross sections, which may help to reduce stochastic noise. The small and well-defined sizes of molecular materials potentially allow small line-edge roughness [2–7].

Tin oxo cages (Fig. 1) are known as negative tone resists and the tin and oxygen atoms in the cage provide a high absorption cross-section of 13 µm-1 at the EUV wavelength (13.5 nm, 92 eV) [8]. In the compound investigated in this work (nicknamed TinOH), the twelve tin atoms in the cage have butyl groups as the organic substituents, and two (OH)-groups are present as counter ions [9,10].

Cardineau et al. [11] were the first to study tin oxo cages as EUV photoresists, and suggested that tin-carbon bond cleavage is the key step in the radiation induced reactions. This model is also supported by computational [12] and experimental studies in more recent work [13–15]. Hinsberg and Meyers [16] proposed a numerical model describing the reaction process in metal oxide-based resists. According to this model, radiation initially leads to the photo-dissociation of ligands from the metal-oxide cores, generating active sites on the metal atoms. The interaction of the active sites with each other leads to cross-linking of the cores through metal oxide binding, finally resulting in the solubility switch of the resist with exposure.

In previous work, TinOH and related compounds were exposed to EUV, Deep UV (DUV), and electrons of various energies [8,12,17–19]. In the work presented here we used broadband extreme ultraviolet (XUV) pulses with energies in the range 25 – 40 eV to induce photo-reactions in thin films of the tin oxo cage resist, and to simultaneously study the transmission changes. The XUV pulses were generated using a tabletop high harmonic generation (HHG) setup developed in our laboratory as the exposure source [20,21]. The recorded absorption changes are quantified, and compared with the predictions from the CXRO database [22] and with EUV results in the literature.

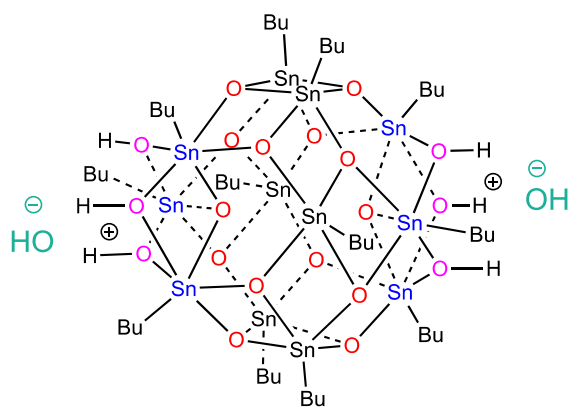

Fig. 1. Tin oxo cage with hydroxide counter ions (TinOH): $[(SnBu)_{12}O_{14}(OH)_6](OH)_2$.

## 2. Experimental
2.1 Sample preparation

TinOH films were spin-coated on free-standing silicon nitride ($Si_3N_4$) membranes (NORCADA; 30 nm thickness) [9,10]. With the same spin coating method film thicknesses of 30 – 40 nm were obtained on other substrates [12,17].

2.2 Experimental set-up and methodology

The general concept of the experiment relies on using a table-top HHG source to simultaneously expose TinOH samples and measure their transmission spectra. By carefully balancing the exposed area and thus the single-shot exposure dose with the collection efficiency of the spectrometer, we can achieve a continuous measurement of the TinOH transmission while simultaneously slowly photo-converting the TinOH cage in the thin film.

Fig. 2 shows a schematic drawing of the HHG set-up used for XUV absorption measurements. Pressures inside the chambers were $10_{-8}$ to $10_{-6}$ mbar. An amplified Ti:Sapphire laser (Solstice Ace; Spectra-Physics) is used for generating extreme ultraviolet pulses by high-harmonic generation. The laser delivers 35 fs, 3.5 mJ pulses at 800 nm center wavelength with a repetition rate of 2 kHz. The fundamental beam is divided into two arms and one beam is frequency-doubled in a BBO crystal. The generated 400 nm beam is recombined with the remaining 800 nm subsequently. The combined beam is focused with a 50 cm focal length concave mirror into a gas cell inside the HHG chamber. An argon pressure of 25 mbar was used in the gas cell for this experiment. The interaction of argon atoms with the high intensity electric field of the laser pulse leads to the generation of high harmonics in the range 25 – 40 eV. The remaining 800 nm light is removed from the XUV pulses with an aluminum filter of 200 nm thickness located between the HHG source and the sample chamber, where the generated harmonics expose the prepared thin films. The XUV pulses were not focused in order to ensure a large illumination area. The broadband transmitted light is directed to the spectrometer chamber where an XUV aberration-corrected concave grating (Hitachi, 1200 lines/mm) disperses the different harmonics onto an XUV charge-coupled device camera (GreatEyes).

The TinOH resist coated on a silicon nitride membrane and an uncoated membrane were mounted next to each other on a motorized stage, and transmission measurements alternated between sample and blank with exposure times of 14.6 seconds.

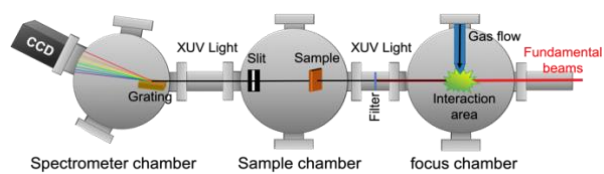

Fig. 2. Schematic of the HHG set-up used for XUV generation. The red line shows the beam path of the fundamental beams and the black path highlights the beam path of the generated XUV pulses.

## 3. Results
A typical high harmonic spectrum in argon generated in our setup is shown in Fig. 3. The energy axis was calibrated using the diffraction equation of the grating. The spectrum shows both odd and even harmonic orders as it was generated with a two-color (800 nm + 400 nm) driving pulse. The spectrum covers 20 – 48 eV, out of which the photon energy range from 25 to 40 eV has sufficiently high flux for measuring transmission changes. For the exposure and the simultaneous transmission measurements the whole HHG spectrum was used. The entire exposure time covers almost 1000 cycles of repeated movements between the resist sample and the uncoated membrane to record the transmitted spectra.

Fig. 4a shows the evolution of the absorbance of the sample ($A$ = -ln $T$) as a function of exposure time. The final exposure time corresponds to an accumulated dose of ~1500 mJ cm$^{-2}$. The reported dose is only a crude estimate, and we are currently working on a dose calibration method to achieve a more precise value. The single-pulse exposure dose was roughly constant over time. In Fig. 4b the average absorbance over the 25 – 40 eV energy window is plotted, with a smooth biexponential trend line. The absorption decreases from the initial $A_{0,av}$ = 1.1 to $A_{\infty,av}$ = 0.73 extrapolated to infinite exposure time.

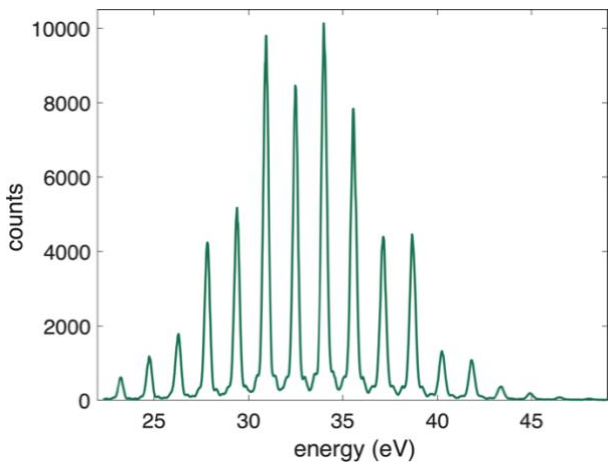

Fig. 3. High-harmonic spectrum generated in argon, recorded after transmission through an Al membrane.

The experimental absorption spectra at the start (dashed red line) and end (dashed blue line) of the exposure time $A_0(E)$ and $A_f(E)$ are shown in Fig. 5, together with the predicted spectra based on the scattering factors [22] at the beginning of exposure $A_{0,pred}$ (red solid line), at the end $A_{f,pred}(E)$ (blue solid line) and for $Sn_{12}O_{18}$ $A_{\infty}(E)$ (black line), which is the hypothetical end product after the dissociation of all butyl groups and the loss of all remaining hydrogen as water.

All spectra show a decrease of absorbance for increasing photon energy, which will be discussed below.

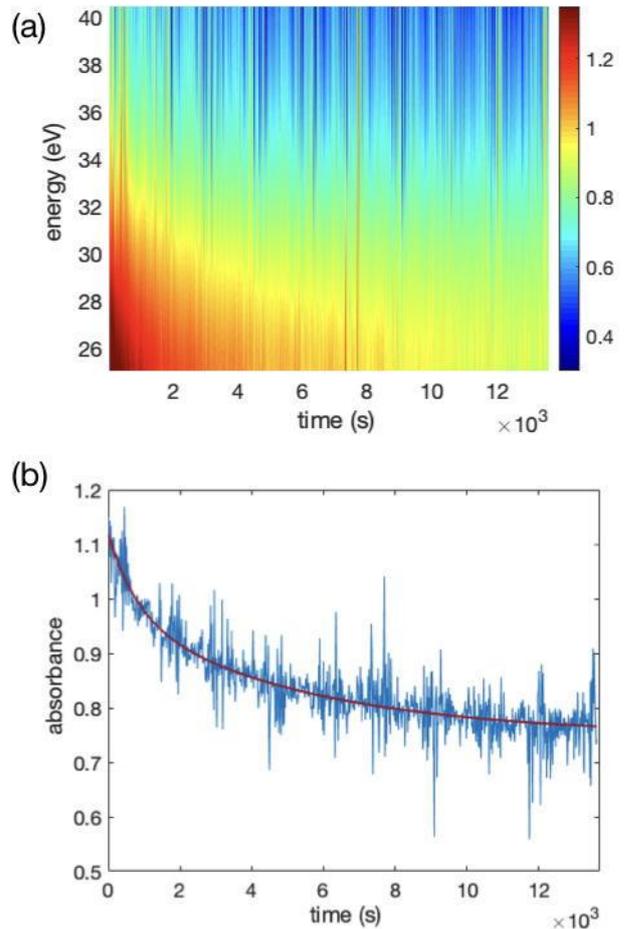

Fig. 4. a) Spectrally resolved absorbance (color coded) exposed up to ~1500 mJ cm$^{-2}$. b) Spectrally averaged absorbance plotted as a function of exposure time. The fit function is a biexponential decay.

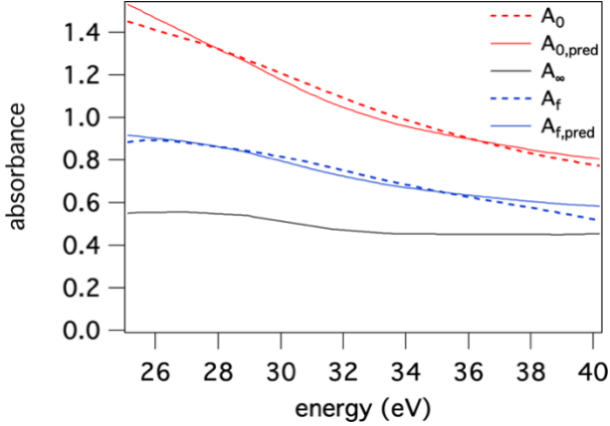

Fig. 5. XUV Absorption spectra of TinOH resist film. Dashed lines are experimental spectra at the beginning (red) and at the end of exposure (blue). The solid lines are calculated from CXRO data [22]. The spectrum $A_\infty$ is the predicted absorbance of $Sn_{12}O_{18}$.

## 4. Discussion

Using the CXRO database values derived from the atomic cross sections $\sigma$ [22], the experimental spectrum at the beginning of exposure can be represented as: $A_0(E) = \sigma_{TinOH}(E) \times S$, where $A_0(E)$ is the experimental absorbance as a function of photon energy as shown in Fig. 5, $\sigma_{TinOH}(E)$ is the predicted absorption cross section of $Sn_{12}C_{48}H_{116}O_{22}$ as a function of energy, and $S$ is the surface density of TinOH cages (in mol cm$^{-2}$) given by $S = \rho z /M$, with $\rho$ being the density of the thin film (in g cm$^{-3}$), $z$ the thickness (in cm) and $M$ the molecular weight (in g mol$^{-1}$).

We obtain $S = 2.61 \times 10^{-9}$ mol cm$^{-2}$, from the measured $A_0(E)$ and predicted $\sigma_{TinOH}(E)$ in a single parameter fit, which results in a good match between the predicted and measured spectra as shown in Fig 5. The fitted value of $S$ corresponds to $z = 34$ nm using $\rho = 1.9$ g cm$^{-3}$, which is a typical density found for various tin cage crystals [9,10,23]. This film thickness is in line with values for the spin-coated films on different substrates[17]. The good match between the predicted and experimental spectra in Fig. 5 also indicates that the tabulated values of the elemental cross sections in the energy range 25 – 40 eV are in good agreement with the experimental data, as was previously found for the EUV energy of 92 eV [8,24].

The absorption decreases with increasing exposure time, which is attributed to the loss of butyl groups. We previously observed the loss of carbon-containing side groups with other types of electron or photon activation in tin oxo cages [13,18]. The hypothetical final product is $Sn_{12}O_{18}$, where all hydrocarbons dissociated ($C_{48}H_{108}$), and all mobile hydrogens are lost as water. In practice, not all butyl group will dissociate. We plot the CXRO predicted spectra for $Sn_{12}O_{18}$ in Fig. 5 as $A_\infty(E)$, using the surface density of tin cages found for the initial film. The final measured spectrum $A_f(E)$ clearly indicates less than complete conversion. By modeling the final experimental spectrum as a linear combination of the initial experimental and predicted final spectra, we estimate the fraction of conversion to be ~60% at the end of the irradiation period in another single-parameter fit, i.e. ~60% of all butyl groups and cleavable water dissociated. The predicted final spectrum is shown as $A_{f,pred}(E)$ in Fig. 5.

The absorption changes of photoresists are usually characterized in terms of the $A$, $B$, and $C$ Dill parameters. To avoid confusion with the symbol for absorbance, we use the symbol $A_{Dill}$. $A_{Dill}$ and $B$ are defined according to equations (1) and (2) [25,26]:

$$A_{Dill} = (A_0 - A_\infty)/z \qquad (1)$$

$$B = A_\infty/z \qquad (2)$$

The Dill $C$ parameter is a measure of the rate of conversion, but due to the uncertainty in the exposure dose value in the present experiment, we do not discuss $C$ in the present work.

$A_{Dill}$, usually given in units of μm$^{-1}$, represents the bleachable part or the exposure dependent absorption of the resist, and $B$ defines the unbleachable part of the absorption. $A_0$ is obtained from the measured initial absorption spectrum, for $A_\infty$ we use the predicted spectrum for complete conversion to $Sn_{12}O_{18}$. The linear absorption coefficient of a material $\alpha$ (μm$^{-1}$) is the sum of the Dill parameters $A_{Dill}$ and $B$. These parameters are shown as a function of energy in Fig. 6. It should be noted that the value of $B$ is purely from the tabulated cross sections, for $\alpha$ and $A_{Dill}$ the values represent the measured spectral shape, but they are scaled using the best fit of the observed spectra to the tabulated cross sections multiplied by the surface density.

The extrapolated change in absorbance averaged over the whole energy range in Fig. 4b suggests that at the end of the exposure time most of the maximum conversion has been reached. This would imply that the final product still contains more carbon and oxygen than $Sn_{12}O_{18}$ (Fig. 5). On the other hand, the gradual decrease in the conversion rate with time renders the extrapolation based on a biexponential decay function uncertain. In our future experiments we plan to shed more light on the dynamics of the photobleaching process.

Fallica and coworkers reported the Dill parameters at 92 eV for a number of photoresists [27], including two tin-based Inpria resists (the structures of which were not disclosed). They showed that the bleachable part $A_{Dill}$ is much smaller than the unbleachable part $B$ at 92 eV. For these two resists, the $B$ parameters were 16 and 19 μm-1 while their A values were ~0.4 and ~0.7 μm-1, respectively For TinOH and related tin oxo cages EUV absorption values around 13 μm-1 were reported [8]. Comparing the present XUV results with the EUV reported values, the unbleachable parts in resists at XUV and EUV energies are similar. In the XUV, the bleachable part of the resist absorption is noticeably larger than at 92 eV.

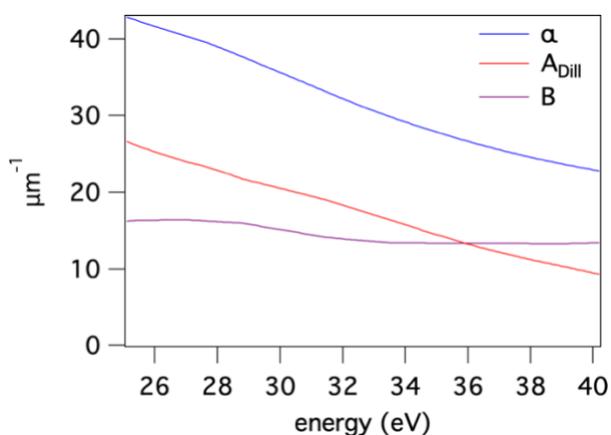

Fig. 6. Dill parameters and linear absorption coefficient for TinOH in the XUV range.

While XUV and EUV radiation have in common that they cause electron ejection as the primary photo-process, the cross sections, and the selectivity for the ionization from different types of orbitals are different. Thus, XUV and EUV irradiation can be complementary tools in EUV photoresist research. The absorption cross-sections of the tin-cage and its constituent elements between 25 and 100 eV are presented in Fig. 7 [22].

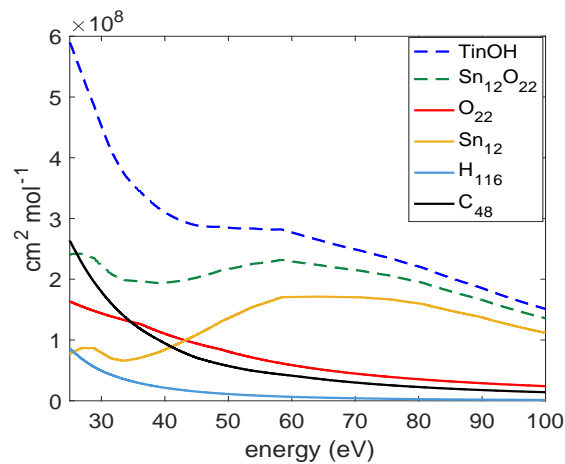

Fig. 7. Photo-absorption cross-sections calculated from the atomic scattering factors from the CXRO database for $C_{48}H_{116}O_{22}Sn_{12}$ (TinOH) and its constituents [22].

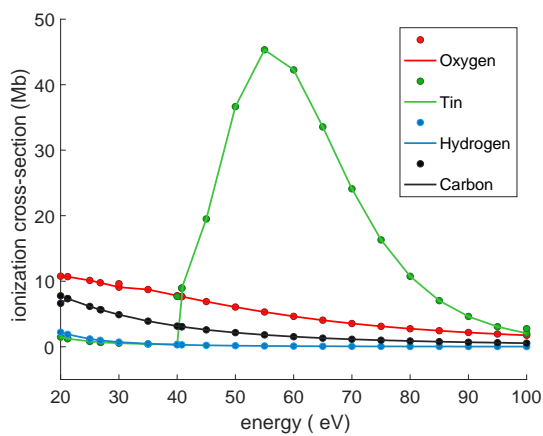

Fig. 8. Photo-ionization cross-section for constituent elements of the resist [28,29].

In the energy range <40 eV the absorption is large, and mostly due to valence ionization of carbon and oxygen. The photo-induced outgassing of the butyl groups and some water molecules from the oxo-metal cage are responsible for the absorbance decrease during exposure in the XUV. The cross sections in Fig. 7 also explain why the absorbance decreases more for low energies during exposure (Figs. 4a and 5), as carbon has a larger contribution to the absorption cross section at lower energies. At 92 eV, however, the absorption is dominated by the Sn atoms, mostly because of ionization from the Sn 4d level. Ionization from these atomic orbitals, with a binding energy of ~25 eV, is energetically feasible in the XUV, but the probability of exciting these 4d electrons is low and increases substantially only above 40 eV [28,30]. This is further illustrated in Fig. 8, in which the computed photoionization cross sections are shown [28]. The large peak at ~60 eV is due to the Sn 4d electrons. Interestingly, recent experimental results indicate that the peak in the formation of Sn 4d photoelectrons occurs near a photon energy of 90 eV, rather than near 60 eV, as was calculated (Fig. 8) [31].

## 5. Conclusion

We used a broadband XUV source from a high-harmonic generation setup spanning from 25 eV to 40 eV to study the photo-induced absorption changes of a tin oxo cage, which is a prominent candidate as photoresist in extreme ultraviolet lithography. We observed a strong absorbance decrease during XUV exposure, that is attributed to XUV-induced tin-carbon bond cleavage. From the absorbance spectra we estimated a ~60% dissociation of all butyl groups for an exposure dose of roughly 1500 mJ cm$^{-2}$.

We extracted the Dill parameters $A_{Dill}$ and $B$, which represent the bleachable and unbleachable part of the absorption. The bleachable part in the studied energy range is larger than in the EUV (92 eV). Compared to EUV exposure, the more pronounced XUV absorbance decrease during exposure was attributed to the photo-induced outgassing of the butyl side groups and some water molecules, which have larger XUV than EUV cross sections due to valence ionization of carbon and oxygen. In contrast, the 4d electrons in tin have a lower cross section in the XUV spectral energy range, but contribute more to the reactions at 92 eV exposure [32].

Our studies introduce table-top high-harmonic sources into the field of photoresist research for EUV lithography. The femtosecond to attosecond pulse durations of HHG sources [33] will make HHG table-top experiments an ideal tool for unraveling the primary steps in the photochemistry of EUV resists, as has been demonstrated on other targets before [34]. Extreme ultraviolet transient spectroscopy has shed light on the coupled electronic and nuclear dynamics in a number of molecules [35], transition metal complexes [36], semiconductors [37,38], and metals [39] in the past. Our work paves the path towards using ultrafast transient extreme ultraviolet spectroscopy in complex inorganic materials in general and photoresists in particular.

## Acknowledgements

This worked was performed in the Advanced Research Center for NanoLithography (ARCNL), a public private partnership of the University of Amsterdam (UvA), the VU University Amsterdam (VU), the Netherlands Organization for Scientific Research (NWO) and the semiconductor equipment manufacturer ASML. We are grateful to Dr. Sandra Mosquera Vazquez, Dr. Robbert Bloem and Sander van Leeuwen for their contributions to the construction of the HHG set-up. This project contributes to the ELENA (Low energy ELEctron driven chemistry for the advantage of emerging NAno-fabrication method) European training network and it has received funding from the European Union's Horizon 2020 research and innovation program under the Marie Skłodowska-Curie grant agreement No 722149. P.M.K. acknowledges support from the Netherlands Organisation for Scientific Research (NWO) for the Veni grant no. 016.Veni.192.254.